# QUANTUM ELECTRIC CIRCUITS ANALOGOUS TO BALLISTIC CONDUCTORS

D. Dragoman – Univ. Bucharest, Physics Dept., P.O. Box MG-11, 077125 Bucharest, Romania

ABSTRACT

The conductance steps in a constricted two-dimensional electron gas and the minimum conductivity in graphene are related to a new uncertainty relation between electric charge and conductance in a quantized electric circuit that mimics the electric transport in mesoscopic systems. This uncertainty relation makes specific use of the discreteness of electric charge. Quantum electric circuits analogous to both constricted two-dimensional electron gas and graphene are introduced. In the latter case a new insight into the origin of minimum conductivity is obtained.

1. INTRODUCTION

The existence of conductance steps equal to $G_0 = e^2/h$ (or $2e^2/h$ if the spin degeneracy is accounted for) is a well-established fact in the physics of the ballistic two-dimensional electron gas (2DEG) subject to a transverse constriction, also called quantum waveguide. In this mesoscopic structure the charge carriers are described by the Schrödinger equation. Despite its universal character for ballistic conductors, the conductance step value is not derived from an uncertainty principle, but from quantum transport considerations in an ideal ballistic conductor subjected to an external electric potential that induces different quasi-Fermi-energies in the reservoirs between which the conductor is sandwiched (see, for example, Ref. 1). Numerous experiments have validated this theoretical result and demonstrated that the conductance varies in steps of $G_0$. On the other hand, charge transport in graphene, in which ballistic charge carriers obey a Dirac-like equation with no effective mass, is characterized by a minimum conductivity value and absence of conductivity steps (for a review, see [2]). This behavior is not fully understood. Present theories (see the bibliographic references in [2]) include but are not limited to chiral-symmetry-preserving disorder [3], Zitterbewegung [4], or maximum of the Fano factor in ideal graphene strips [5], no agreement being possible between theories based on such different assumptions.

The purpose of this paper is to show that the stepwise variation of the conductance in ballistic systems in which charge carriers obey the Schrödinger equation, as well as the minimum conductivity value and the apparent absence of conductivity steps in graphene, can be better understood by quantizing electric circuits that mimic their behavior. In particular, the step-like character of the conductance can be obtained from a quantization of the Ohm law in an electric circuit that contains a biased conductance. The value of the conductance steps is thus associated with a quantum uncertainty relation between current and conductance in this circuit. On the other hand, the transport of charge carriers in graphene can be mimicked by the quantization of an electric circuit containing a biased conductance and a current generator. The physical meaning of the circuit components and the reason why the quantization of electric circuits can describe the behavior of charge transport in various ballistic conductors is discussed.

## 2. THE QUANTUM ELECTRIC CIRCUIT ANALOGOUS TO A CONSTRICTED TWO-DIMENSIONAL ELECTRON GAS

It is natural to assume that the electrical circuit equivalent to a ballistic conductor is just a biased conductance, as illustrated in Fig. 1. The classical Ohm law is then:

$$I = dq/dt = GV \qquad (1)$$

where $I$ is the current, $G$ the macroscopic conductance, $V$ the applied potential and $q$ the electrical charge. Equation (1) can be quantized by duly taking into account the discreteness of the electric charge, in a manner similar to that in Ref. 6. The quantization of an $LC$ electric circuit in Ref. 6 proceeded by finding first a Hamiltonian in the conjugate variables electric charge and (scaled) electric current. In our case the conjugate variables are the electric charge $q$ and the conductance $G$, the corresponding classical Hamiltonian from which (1) can readily be obtained being $H(q,G) = VG^2/2$. Contrary to the Hamiltonian in Ref. 6, the dimension of $H(q,G)$ does not suggest a straightforward relation to the energy; such Hamiltonians are not uncommon in classical mechanics and, in particular, in classical optics [7]. One could recover the dimension of energy normalized to the elementary electron charge $e$ if $G$ would be adimensional (properly scaled).

The classical Hamiltonian $H(q,G)$ is next quantified by introducing operators of electric charge and conductance, $\hat{q}$ and $\hat{G}$, respectively, that satisfy the commutation relation

$$[\hat{q}, \hat{G}] = i\beta \tag{2}$$

with $\beta$ a positive constant. There is no reason to set $\beta$ equal to $\hbar$, but a non-zero $\beta$ value is required to pass from a classical to a quantum treatment of the electrical circuit under consideration. In order to recover the (scaled to $e$) energy meaning of the Hamiltonian eigenvalues, we take the quantum Hamiltonian to be

$$\hat{H} = V(e\hat{G}/\beta)^2/2, \tag{3}$$

the operator $e\hat{G}/\beta$ having adimensional eigenvalues. Note that this Hamiltonian is identical in form to that of a free quantum particle, the difference being that the charge operator (which corresponds to the position operator for the free particle) has now discrete values; the operator $\hat{G}$ corresponds to the momentum operator for the free particle.

As in Ref. 6 we impose that the eigenvalues of $\hat{q}$ are discrete, i.e. that

$$\hat{q}|n\rangle = ne|n\rangle, \tag{4}$$

with $n$ an integer. By denoting with $|n\rangle$ the eigenfunctions of the charge operator we implicitly assume that they span a Fock space with ladder operators $\hat{Q} = \exp(ie\hat{G}/\beta)$ and $\hat{Q}^+$. The eigenfunctions of the operator $\hat{G}$ can then be expressed as

$$|G\rangle = \sum_{n \in Z} a_n \exp(ineG/\beta)|n\rangle \tag{5}$$

with suitably chosen coefficients $a_n$ [6].

To properly account for the discreteness of the electric charge, we follow the mathematical treatment in [6] and define right and left discrete derivative operators as $\nabla_e^r = (\hat{Q}-1)/e$ and $\nabla_e^l = (1-\hat{Q}^+)/e$, respectively, their action on an arbitrary function $f(n)$ being

$$\nabla_e^r f(n) = [f(n+1) - f(n)]/e, \quad \nabla_e^l f(n) = [f(n) - f(n-1)]/e. \tag{6}$$

In terms of these operators the time-independent Schrödinger equation becomes

$$\left[-\frac{eV}{2}(\nabla_e^r - \nabla_e^l)\right]|\Psi\rangle = \left[-\frac{V}{2}(\hat{Q} + \hat{Q}^+ - 2)\right]|\Psi\rangle = (E/e)|\Psi\rangle. \tag{7}$$

Taking into consideration also that $\hat{Q}|G\rangle = \exp(ieG/\beta)|G\rangle$, it is finally found that the corresponding finite-difference Schrödinger equation in the $G$ representation is

$$-eV[\cos(eG/\beta) - 1]\Psi(G) = E\Psi(G). \tag{8}$$

This equation is satisfied if $eG/\beta = \text{Arccos}(1 - E/eV) + 2m\pi$ with $0 \leq E/eV \leq 2$ and $m$ integer, which implies that $G$ is multivalued for a given $E$ and $V$, and that $G$ varies in steps of $2\pi\beta/e$. A more precise statement can be made if we interpret $E$ as the excess energy in the circuit. In this case, if all the energy provided by the electric source falls on the conductance, we have $E = 0$ and

$$eG/\beta = 2m\pi. \tag{9}$$

It then follows that the conductance varies in steps of $2\pi\beta/e$. The integer $m$ in (9) should, however, have non-negative values since a negative value is not compatible with a passive electrical device. This astonishing result is obtained by quantizing the macroscopic Ohm law, without any regard to the quantum motion of electrons; only the discreteness of the electron charge has been used via relations (4)-(9). In order to recover the known result that the conductance steps are equal to $G_0 = e^2/h$, one should set $\beta = e^3/2\pi\hbar$. Spin degeneracy can then be incorporated in the theory by simply multiplying with 2 the value of $G_0 = e^2/h$.

For the quantized electric circuit considered above, the correspondence with the classical theory does not follow from $\hbar = h/2\pi \to 0$ as in usual quantum mechanics, but from $e \to 0$ for a finite value of $\hbar$. This suggests that a full quantum mechanical treatment of charged particles should be governed by two finite parameters: $e$ and $\hbar$, instead of only $\hbar$. In particular, a proper consideration of the discrete nature of $e$ leads to the recovery of the minimum conductance value from a new uncertainty relation rather than from the conventional derivation based on the Schrödinger equation for ballistic electrons.

It is important to understand why this very simple model of a biased conductance reproduces the conductance steps of a constricted 2DEG. If the electrons in the conduction band of a semiconductor are confined in the $(x,y)$ plane by a potential $U(z)$ so that only the lowest subband corresponding to $z$ confinement is occupied, the actual free 2DEG is described by the time-independent Schrödinger equation

$$-\frac{\hbar^2}{2m}\left(\frac{\partial^2}{\partial x^2}+\frac{\partial^2}{\partial y^2}\right)\Psi(x,y) = (E-E_0)\Psi(x,y) \tag{10}$$

where $m$ is the effective mass of the electrons in the conduction band and $E_0$ encompasses both the conduction band edge and the cut-off energy of the occupied $z$-confined subband. Let us now suppose that this 2DEG is further confined along $y$ by an applied negative voltage, as in a split-gate configuration. Then, the energy dispersion relation of electrons in the quantum waveguide is

$$E(n,k) = E_0 + \hbar^2 k^2/2m + \varepsilon_n \tag{11}$$

where $k$ is the wavevector component along the propagation direction (along $x$) and $\varepsilon_n$ are the eigenenergies of the confined motion along $y$. Each $n$ defines a subband. If spin degeneracy is taken into account, the current of free electrons (with quantum transmission $T(E)=1$) through the constriction in the presence of an applied bias is given at low temperatures by

$$I = \left(\frac{2e}{2\pi}\right)\sum_n \int_{\mu_1}^{\mu_2} dE \left(\frac{dk}{dE}\right) v(n,k) = V\sum_n (2e^2/h) = VN(2e^2/h) \qquad (12)$$

where the sum is taken over all $N$ occupied subbands, for which $\varepsilon_n < \hbar^2 k_F^2/2m$ with $k_F$ the Fermi wavenumber, and the bias $V$ is applied between the electron reservoirs that surround the ballistic conductor and have different quasi-Fermi energies $\mu_1$, $\mu_2$, such that $eV = \mu_1 - \mu_2$. The expression for the current in (12) is independent of the dispersion relation since the electron velocity is defined as $v(n,k) = \hbar^{-1} \partial E(n,k)/\partial k = \hbar k/m$. The total conductance $NG_0$ can be modified in experiments since the number of occupied subbands $N$ is varied by the transverse confining potential in the split-gate configuration.

From (3) and (10) it follows that, indeed, the conductance is analogous to the momentum operator for ballistic electrons, and that the $m$ integer in (9) corresponds to the number of occupied subbands; whereas the latter parameter can be related to the transverse confining potential, $m$ seems to be arbitrary. The discreteness of $m$ is related to the discreteness of electric charge in the quantized electric circuit, whereas for ballistic electrons the (discrete) occupied subbands arise from a transverse boundary condition. (A finite number of occupied subbands should correspond to a finite number of electrons in the system but a finite sum in (5) leads to non-orthogonal and non-complete eigenfunctions of the conductance operator.) Although rude as a model and based on different physical considerations, the quantization of the electric circuit introduces the correct conductance step, i.e. assigns a conductivity equal to $G_0$ per carrier, but fails to indicate (at least in the simplified case presented in this section) the exact value of the conductance (the exact number of steps). The reason is that the transverse boundary conditions, which are related to external potentials and hence prone to experimental manipulation, correspond to the (fixed) charge unit *e*, which cannot be experimentally varied.

3. THE QUANTUM ELECTRIC CIRCUIT ANALOGOUS TO GRAPHENE

Stimulated by the discussion in the previous section, one may wonder if there is an electric circuit that mimics the behavior of conductivity in graphene. More precisely, in this case the conductivity should not display steps and it should always be finite (the $G = 0$ value, possible for a quantum waveguide discussed in the previous section, does not occur in graphene). Experiments have shown that the minimum conductivity in graphene is around $4e^2/h$ in zero-field conditions (near the Dirac point), where the carrier concentrations should vanish [8].

One may suppose that, since we refer to conductivity measurements, the classical circuit we are searching for should include a biased conductance; the electric circuit we are considering in this section is represented in Fig. 2 and is described classically by the equation

$$I = dq/dt = GV - I_0. \qquad (13)$$

This electric circuit should be quantized differently than for the biased conductance case, i.e. assuming a different Hamiltonian and spinor-like states, suitable for Dirac-like charge carriers because the charge carriers in graphene behave as massless Dirac-like fermions [2], the electrons and holes corresponding, respectively, to positive and negative energy solutions of the linear dispersion relation.

More precisely, we introduce operators of electric charge and conductance that satisfy a commutation relation similar to (2) in which, in order to distinguish between the two situations, we replace $\beta$ with $\beta'$. Then, inspired by the transition from the time-independent non-relativistic Schrödinger equation for a free particle $(\hat{p}^2/2m - E)|\Psi\rangle = 0$ to the time-indepen-dent relativistic Dirac equation in quantum mechanics $(\sigma_z \cdot pc + \sigma_y mc^2 - EI_2)|\Psi\rangle = 0$, with $\sigma_{z,y}$ spin matrices and $I_2$ the two-dimensional unit matrix, we suppose that the Dirac-like equation of the electric circuit that should be used in the case of graphene is

$$[V\sigma_z \cdot (e\hat{G}/\beta') - \sigma_y(I_0 e/\beta') - (E/e)I_2]|\Psi\rangle = 0 \qquad (14)$$

Note that in the above equation the intensity of the current source, $I_0$, has been normalized to $\beta'/e$ for dimensional considerations and that it corresponds to an effective-mass for the ballistic electrons. Note that this "effective-mass" does not appear in the Dirac-like equation in which the non-commuting operators are those associated to position and momentum; the meaning of this parameter is discussed later on. The quantum state in (14) is a spinor, its two components describing the two chiral charge carriers in graphene [2].

In a similar manner as in the previous section, and inspired from Ref. 6, one can write (14) as

$$\begin{aligned}&\left[V\sigma_z \cdot \frac{e}{2i}(\nabla_e^r + \nabla_e^l) - \sigma_y(I_0 e/\beta') - (E/e)I_2\right]|\Psi\rangle \\ &= \left[V\sigma_z \cdot \frac{1}{2i}(\hat{Q} - \hat{Q}^+) - \sigma_y(I_0 e/\beta') - (E/e)I_2\right]|\Psi\rangle = 0\end{aligned}, \qquad (15)$$

the corresponding Dirac-like equation in the $G$ representation being

$$[V\sigma_z \cdot \sin(eG/\beta') - \sigma_y(I_0 e/\beta') - (E/e)I_2]\Psi(G) = 0. \qquad (16)$$

In the same conditions as in the previous section, i.e. for $E = 0$, equation (16) is satisfied if

$$eG/\beta' = \text{Arcsin}(I_0 e/\beta' V) + 2m\pi \qquad (17)$$

with $m$ a non-negative integer and $|I_0 e / \beta' V| \leq 1$. If we assume $\beta' = \beta$ we find that, again, $G = m e^2 / h$, i.e. the conductivity varies in steps of $e^2 / h$. This conclusion is supported by the fact that the conductance value as derived from (12) is independent of the dispersion relation.

However, this is not what experimental observations suggest. The argument that the theory developed in the previous section does not apply in the case of graphene, since the latter is a two-dimensional crystal and not a quantum waveguide, is not valid. The experiments on the graphene conductivity have been performed on a multi-terminal Hall bar configuration, which can be considered as a constricted graphene structure in a similar way as the split-gate configuration is a constricted 2DEG. Moreover, in ideal conditions the conductance measurements in a multi-terminal Hall bar configuration gives the same result as a two-terminal experiment.

To explain why no conductivity steps are observed in graphene, we must observe first that the dispersion relation in graphene $E = \pm |\hbar v_F k|$, with $v_F$ the Fermi velocity, is not the $m \to 0$ limit of the Dirac equation of a particle with mass, which would look like $E = \pm |\hbar c k|$, but the $m \to 0$ limit of a scaled Dirac equation. More precisely, if the energy $E$ and the wavenumber $k$ in graphene retain their meaning and definition as in a common semiconductor, the uncertainty relation between position and momentum operators in graphene is no longer $[\hat{x}, \hat{p}] = i\hbar$, but $[\hat{x}, \hat{p}] = i\hbar (v_F / c)$. (Note that the uncertainty relation between position and momentum operators is the same for the Schrödinger and the Dirac equations for a particle with mass.) Since the charge and conductance operators introduced in the previous section are analogous to the position and momentum operators, respectively, we can assume that for graphene $\beta' = (v_F / c) \beta$, which imply that the conductance steps in graphene are scaled with $v_F / c$ with respect to those in a quantum waveguide. These steps are about 300 times smaller than $G_0$ and, therefore, not observable.

The minimum conductivity, equal to $G_0 = e^2/h$ per spin per valley (the spin and valley degeneracy introduce a multiplication factor of 4), is then recovered if we set $I_0/V = (\beta'/e)\sin(G_0 e/\beta')$. Note that $I_0$ depends on the "velocity mismatch" parameter $v_F/c$. The current generator signifies, in our opinion, the energy necessary for the creation of a one-charge state. Contrary to common semiconductors, where independent electron and hole states exist prior to the application of a bias, the applied voltage only setting them in controlled motion, electron and hole states in graphene do not exist separately and must first be created. Due to the discreteness of the electronic charge, the zero-charge state $|n\rangle$ with eigenvalue $n = 0$, which is an ambiguous charge state since it does not represent neither an electron or a hole, and the one-charge state with $n = 1$ that represents either an electron or a hole are distinct and one cannot obtain the first by continuously approaching the zero limit of energy in the energy dispersion relation. The energy needed for the creation of the one-charge state that is required for the measurement of conductivity is embodied in the parameter $I_0$. In the Dirac-like equation (14) it appears as an effective-mass parameter, which represents the quantum circuits' "inertia" to controlled movement characterized by the conductance operator. This "inertia" expresses the energy necessary in graphene to generate un-ambiguous quantum charge states.

Note that this interpretation of the minimum conductivity is intrinsically linked with the discreteness of electric charge, which is not assumed in the Dirac equation of charged particle with or without mass. The explicit consideration of this fact leads to a new understanding of the origin of the minimum conductivity in graphene. Moreover, the arguments and mathematical treatment presented in this section, can be extended also to bilayer graphene, which is characterized by the same minimum conductivity value per spin per valley. This recently confirmed experimental fact was characterized as "challenge for theory" [9]; our approach can accommodate the minimum conductivity value for the bilayer graphene by explicitly accounting for the discreteness of the electric charge.

4. CONCLUSIONS

In conclusion, the conductivity step value for ballistic Schrödinger-like charge carriers and the minimum conductivity and apparent absence of conductivity steps for Dirac-like charge carriers have been derived from a new uncertainty relation between the electric charge and conductance/conductivity in electrical circuits. This uncertainty relation exploits the discrete nature of the electric charge and should supplement the Heisenberg-type uncertainty for a complete description of charged particles. Moreover, the quantized electrical circuits that mimic the conductance behavior of ballistic electrons offer, especially for the case of graphene, a new insight into the origin of the minimum conductivity.

REFERENCES

[1]     D.K. Ferry, S.M. Goodnick, *Transport in Nanostructures*, Cambridge Univ. Press, 1997

[2]     A.K. Geim, K.S. Novoselov, "The rise of graphene", Nature Materials 6, 183-191, 2007

[3]     P.M. Ostrovsky, I.V. Gornyi, A.D. Mirlin, "Electron transport in disordered graphene", Phys. Rev. B 74, 235443, 2006

[4]     M.I. Katsnelson, "Zitterbewegung, chirality, and minimal conductivity in graphene", Eur. Phys. J. B 51, 157-160, 2006

[5]     J. Tworzydło, B. Trauzettel, M. Titov, A. Rycerz, C.W.J. Beenakker, "Sub-Poissonian shot noise in graphene", Phys. Rev. Lett. 96, 246802, 2006

[6]     Y.-Q. Li, B. Chen, "Quantum theory for mesoscopic electric circuits", Phys. Rev. B 53, 4027-4032, 1996

[7]     R.K. Luneburg, *Mathematical Theory of Optics*, Univ. California Press, Berkeley, 1964

[8]     K.S. Novoselov, A.K. Geim, S.V. Morozov, D. Jiang, M.I. Katsnelson, I.V. Grigorieva, S.V. Dubonos, A.A. Firsov, "Two-dimensional gas of massless Dirac fermions in graphene", Nature 438, 197-200, 2005

[9]     K.S. Novoselov, E. McCann, S.V. Morozov, V.I. Fal'ko, M.I. Katsnelson, U. Zeitler, D. Jiang, F. Schedin, A.K. Geim, "Unconventional quantum Hall effect and Berry's phase of $2\pi$ in bilayer graphene", Nature Physics 2, 177-180, 2006

FIGURE CAPTIONS

Fig. 1  The electric circuit that mimics, upon quantization, the charge transport in a two-dimensional electron gas.

Fig. 2  The electric circuit that mimics, upon quantization, the charge transport in graphene.

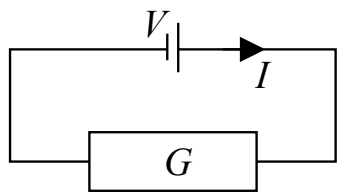

Fig. 1

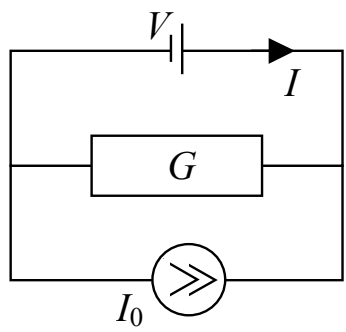

Fig. 2